\documentclass[twocolumn,showpacs,preprintnumbers,amsmath,amssymb]{revtex4}

\usepackage{graphicx}
\usepackage{dcolumn}
\usepackage{bm}
\usepackage{color}

\def\add#1{{{}{#1}}}          

\begin{document}

\title{Shell to shell energy transfer in MHD,\\
       Part II: Kinematic dynamo}

\author{Pablo D. Mininni}
    \email{mininni@ucar.edu}
\author{Alexandros Alexakis}
    \email{alexakis@ucar.edu}
\author{Annick Pouquet}
\affiliation{National Center for Atmospheric Research,
             P.O. Box 3000, Boulder, Colorado 80307}

\date{\today}

\begin{abstract}

We study the transfer of energy between different scales for forced 
three-dimensional MHD turbulent flows in the kinematic dynamo regime.
Two different forces are examined:  a non-helical Taylor Green flow with
magnetic Prandtl number $P_M=0.4$, and a helical ABC flow with $P_M=1$.
This analysis allows us to examine which scales of the velocity flow
are responsible for dynamo action, and \add{identify} which scales of the magnetic field
receive energy directly from the velocity field and which scales receive
magnetic energy through the cascade of the magnetic field from large to
small scales. Our results show that the turbulent velocity fluctuations   
are responsible for the magnetic field amplification in the small scales 
(small scale dynamo) while the large scale field is amplified mostly due
to the large scale flow. A direct cascade of the magnetic field energy from
large to small scales is also present and is a complementary mechanism for the
increase of the magnetic field in the small scales.
Input of energy from the velocity field in the small magnetic scales
dominates over the energy that is cascaded down from the large scales
until the large-scale peak of the magnetic energy spectrum is reached.
At even smaller scales, most of the magnetic energy input is from the cascading process.

\end{abstract}

\maketitle


\section{ \label{Intro} Introduction }

Dynamo action is often invoked to explain the generation and
sustainment of magnetic fields in astronomical objects. In the
magnetohydrodynamic (MHD) dynamo, an initially small magnetic field
is amplified by currents induced solely by the motion of a
conducting fluid \cite{Moffatt}. In typical astrophysical
situations where amplified magnetic fields are met, the velocity
field is composed of a large scale flow (e.g. rotation and/or
meridional flows) together with turbulent fluctuations in smaller
scales. As an example, in the sun both large and small scale magnetic fields
are observed. The large scale components \add{of the magnetic field} 
are generated by a large-scale flow \cite{Dikpati99}. 
Simulations also show that the small
scale magnetic fields can be generated by turbulent fluctuations in
the convective region \cite{Cattaneo99}.
Understanding the generation of magnetic fields under these conditions and
the role played by the two components of the flow
(large scale and turbulent) is today a crucial aspect of dynamo theory. 

Dynamos are often classified as small-scale and large-scale
dynamos, depending on the properties of the amplified magnetic field
\cite{Vainshtein72}. In large-scale dynamos, the focus is on whether a
flow can amplify and sustain magnetic fields at
scales larger than the velocity integral scale. This interest is motivated by
astrophysical problems where large scale magnetic fields are
actually observed, such as the dipolar component in stars and
planets. The amplification of the magnetic field in these scales 
is usually explained by invoking a turbulent $\alpha$-effect and/or 
amplification due to a large scale flow. The linear (or kinematic) regime 
of large-scale dynamo action has been studied with the use of mean field 
theory \cite{Steenbeck66,Krause}, MHD closures \cite{Pouquet76}, and with 
the aid of numerous direct numerical simulations (DNS) (see {\it e.g.} 
\cite{Meneguzzi81,Brandenburg01,Gomez04}).
In theoretical investigations of large-scale dynamo 
action, helical flows are generally considered, that are thought of as better
candidates for amplifying the magnetic field at larger scales.
However the presence of helicity is not necessarily needed to
generate large scale magnetic fields \cite{Gilbert88}; 
they can also be amplified in non-helical flows if
anisotropy \cite{Nore97}, or other mean field effects 
\cite{Urpin02,Geppert02} are present. 


Small-scale dynamos on the other hand amplify magnetic fields on scales
smaller than the energy containing scales of the turbulence
\citep{Kazantsev68,Zeldovich,Schekochihin02,Haugen03,Schekochihin04b}.
Theoretical investigations usually involve assumptions of non-helical
velocity fields, $\delta$-correlated in time (as a simplifying
approximation to a turbulent flow), and often the limit of
large magnetic Prandtl number $P_M>1$ is considered. 
Numerically small-scale dynamos have been investigated in Refs. 
\cite{Meneguzzi81,Haugen03,Schekochihin04b}. 
Here we note that an argument due to Batchelor
\cite{Batchelor50} suggests that this dynamo can only operate if
$P_M>1$. However, there are reasons to believe the small-scale
dynamo can work even when $P_M<1$ if the magnetic Reynolds number
$R_M$ is large enough \cite{Schekochihin04a,Ponty05}.

However, this separation between large-scale and small-scale dynamos
is in some cases artificial and may be misleading. Most astronomical objects display
a large scale flow with turbulent fluctuations at smaller scales, 
and both large and small scales magnetic fields are observed. 
The transition between the two magnetic fields is often smooth 
and a clear distinction between the two cannot be made. 
This has led some authors to develop models trying to unify the
two regimes \cite{Subramanian99}. 
Furthermore, the two amplification mechanisms
in the small and large scales are coupled in many cases 
and cannot be considered independently.
According to mean field theory \cite{Krause,Zeldovich}, 
the large scale magnetic field in
a turbulent dynamo results from the small scale (helical) velocity fields.
Moreover, concerning the amplification of small scale magnetic fields, 
it has been argued that when a large scale magnetic field is present, 
small scales can be generated
by the distortion of large scale field lines (see e.g. \cite{Moffatt}), 
even in the absence of self-excitation (small-scale dynamo action).
This is a common assumption in mean field dynamos, where it is 
often considered that the needed small scale magnetic fields are only 
fed by the large scale field through a nonlinear cascade process.

In the presence of both a large scale flow and turbulent
fluctuations, the role played by the different scales involved in
the amplification process is thus of crucial importance and 
is not well understood. When magnetic fields are present at scales 
both smaller and larger than the energy containing scales of the 
velocity field, it is not clear what portion of the small scale magnetic 
field is generated by direct cascade of magnetic energy from the large scales, 
and what from self-excitation. 
Furthermore it is not well understood what portion of the 
amplification of the large or small scale dynamo is due to the 
forced \add{component of the} flow and what part is due to the turbulent fluctuations which emerge through nonlinear interactions at high Reynolds number.
To answer these questions, a detailed study of the energy transfer
from the different velocity scales to the different magnetic scales 
is required. This kind of approach naturally raises the question of 
the locality (in Fourier space) of the interactions that are taking 
place in a turbulent dynamo.

In a companion paper \cite{Alexakis05} (hereafter referred to as Paper I),
the transfer of energy between the velocity and magnetic
field at different scales was studied for mechanically forced MHD turbulence
in a steady state where both fields are in quasi equipartition,
by introducing the energy transfer functions
between different shells of wavenumbers in Fourier space.
In this paper, we present shell-to-shell
energy transfers during the kinematic regime of two different MHD dynamos.
Our main interest is to identify which velocity field scales are 
responsible for the amplification of the large and small scale 
magnetic field, which scales of the magnetic field
receive most of the energy, and how the magnetic energy cascades 
among the different scales.

In Sec. \ref{Transfer} we present a brief review of the equations
and definition of transfer functions needed to study this problem, and
in Sec. \ref{Results} we give the results from simulations; we also
discuss in this section some details of the nonlinear saturation of the
dynamo. Finally, in Sec. \ref{Discussion} we present the conclusions of
our work.

\section{ \label{Transfer} The transfer functions}

We will consider the incompressible MHD equations,
\begin{equation}
\partial_t {\bf u} + {\bf u}\cdot \nabla {\bf u} = - \nabla p +
    {\bf b}\cdot \nabla {\bf b} + \nu \nabla^2 {\bf u} +{\bf f}
\label{eq:momentum}
\end{equation}
\begin{equation}
\partial_t {\bf b} + {\bf u}\cdot \nabla {\bf b} =
    {\bf b}\cdot \nabla {\bf u} + \eta \nabla^2 {\bf b} ,
\label{eq:induc}
\end{equation}
where ${\bf u}$ is the velocity field, ${\bf b}$ is the magnetic
field, $\nu$ is the kinematic viscosity, $\eta$ is the magnetic
diffusivity, $p$ is the total pressure and ${\bf f}$ a constant external force. This equations
are accompanied by the conditions
$\nabla \cdot {\bf u} = 0 = \nabla \cdot {\bf b}$. Equations
(\ref{eq:momentum}) and (\ref{eq:induc}) are solved in a periodic
domain using a pseudospectral method with the $2/3$ dealiasing
rule and second order Runge-Kutta to advance in time.

We are interested in the kinematic regime of the dynamo, where a
small magnetic seed is amplified exponentially without modifying the
velocity field (i.e., the effect of the Lorentz force on the
velocity field is negligible). 

To this end, we made two numerical
simulations using a grid of $256^3$ points under the following
procedure. First, a hydrodynamic simulation was performed to obtain
a turbulent steady state. 
Then, a random small magnetic field was
introduced and the simulation was carried to observe exponential
amplification of the magnetic energy. The data were analyzed
during this stage and as the systems approached saturation.

Two expressions for the external force were used:
Taylor-Green (hereafter called TG), and ABC. The TG forcing
is non-helical (${\bf f \cdot \nabla \times f}=0$ pointwise), while 
the ABC forcing is of maximum helicity and the resulting flow has 
non-negligible helicity (for a description of the resulting flows see 
e.g. \cite{Mininni05a,Mininni05b}). 
In both simulations, the amplitude of the external force was set to
obtain a unity r.m.s. velocity, and the characteristic wavenumber of
the force was chosen to obtain a large scale flow at $k_F \sim 3$.
The TG simulation had $\nu = 2 \times 10^{-3}$ and $\eta = 5 \times
10^{-3}$ (the magnetic Prandtl number in this simulation was $P_M =
\nu/\eta = 0.4$). In the ABC run, $\nu = \eta = 2 \times 10^{-3}$
($P_M = 1$). \add{The mechanical Reynolds numbers reached by the two flows 
are $Re=675$ for the Taylor Green flow and $Re=820$ for the ABC}
\cite{Mininni05a,Mininni05b}.

As we stated in the introduction, we are interested in
quantifying the rate of energy transfer from the different
scales of the velocity field to the different scales of the magnetic
field. To rigorously  define the velocity and magnetic field at
different scales we introduce the shell filtered
velocity and magnetic field components ${\bf u}_K({\bf x})$ and 
${\bf b}_K({\bf x})$,
where the subscript $K$ indicates that the field has been filtered to
keep only the modes in the Fourier shell $[K,K+1)$ (hereafter called
the shell $K$). Clearly the sum of all the $K$ components gives back
the original field. We are interested therefore in the rate that 
energy from the velocity or magnetic field at a given shell $Q$ 
is transferred into kinetic or magnetic energy at another shell $K$. 
From the MHD equations, 
by doting Eq. (\ref{eq:induc}) with ${\bf b}_K$ and integrating over 
space, we obtain the evolution of the magnetic 
energy $E_b(K)=\int b_K^2/2 \, dx^3$ in the shell $K$:
\begin{equation}
\partial_t { E}_{b}(K) = 
\sum_Q [{\mathcal T}_{ub}(Q,K)+{\mathcal T}_{bb}(Q,K)] - 
\eta {\mathcal D}_b(K),
\label{eq:Eb}
\end{equation}
where we have introduced the two transfer functions ${\mathcal T}_{ub}(Q,K)$ 
and ${\mathcal T}_{bb}(Q,K)$ as defined below.
The transfer rate of kinetic energy 
at the shell $Q$ into magnetic energy at the shell $K$ is defined as:
\begin{equation}
{\mathcal T}_{ub}(Q,K) = \int{ {\bf b}_K ({\bf b}\cdot\nabla){\bf
u}_Q
    d{\bf x}^3} , \label{eq:Tub}
\end{equation}
and the transfer rate of magnetic energy
from the shell $Q$ into the shell $K$ is defined as:
\begin{equation}
{\mathcal T}_{bb}(Q,K) = -\int{ {\bf b}_K ({\bf u}\cdot\nabla){\bf b}_Q
    d{\bf x}^3} . \label{eq:Tbb}
\end{equation}
The transfer ${\mathcal T}_{ub}(Q,K)$ is due to the stretching of magnetic 
field lines by the velocity field gradients and leads to 
energy input in the magnetic field.  This term is responsible for dynamo 
action, i.e. conversion of kinetic energy into magnetic energy.
The function ${\mathcal T}_{bb}(Q,K)$ is due to the advection of
magnetic field \add{vector components} by the velocity field and it does not amplify
the total magnetic energy. Instead, it is responsible for the
redistribution of magnetic energy among the different shells
and it is related with the cascade of magnetic energy from larger to 
smaller scales. Finally, we introduced the dissipation rate 
${\mathcal D}_b(K)$ in the shell $K$ defined as:
\[{\mathcal D}_b(K)=-\int |\nabla \times {\bf b}_K|^2 d{\bf x}. \]
More  detailed definitions of these transfer terms and their general 
properties can be found in Paper I.

We measured the transfer functions based on Eqs. 
(\ref{eq:Tub}) and (\ref{eq:Tbb}) using ten different outputs 
for each run during the kinematic regime. 
The transfers were normalized using the total magnetic energy,
and were then averaged. As the system was approaching saturation
and was deviating from the exponential growth, single time
outputs were used and the transfer functions were normalized
using the total magnetic energy but were not averaged, since in this
stage the normalized magnetic energy spectrum is changing with time.  
From here on, we will use the notations ${\mathcal T}_{ub}(Q,K)$ and
${\mathcal T}_{bb}(Q,K)$ for the normalized transfer functions 
${\mathcal T}_{ub}(Q,K)/\sum_{K'}E_b(K')$ and
${\mathcal T}_{bb}(Q,K)/\sum_{K'}E_b(K')$, unless otherwise noted.

\section{ \label{Results} Results}

\subsection{ \label{kinematic} The kinematic regime}

\begin{figure}
\includegraphics[width=8cm]{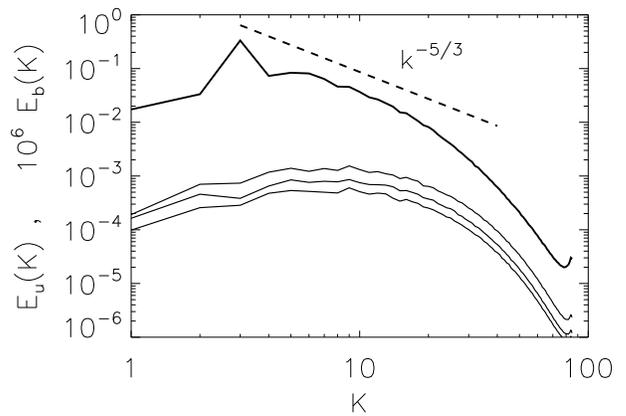}
\caption{\label{fig_01} Spectra of kinetic energy (thick solid line)
and magnetic energy (thin solid line) \add{scaled up by a factor $10^6$}
for the Taylor Green runs during 
the kinematic dynamo regime. The dashed line indicates the Kolmogorov 
slope as a reference. Note that during this stage, all the magnetic 
modes grow with approximately the same rate.}
\end{figure}
\begin{figure}
\includegraphics[width=8cm]{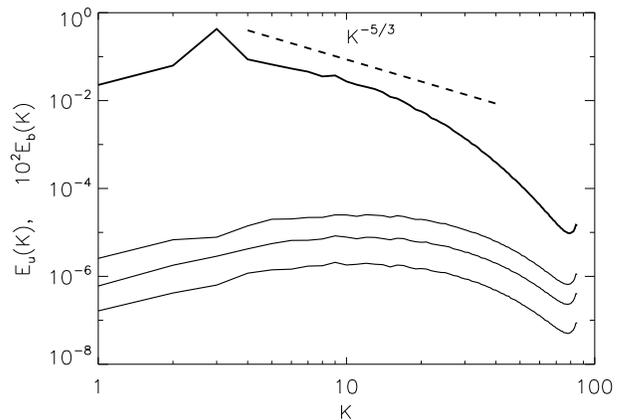}
\caption{\label{fig_02} Spectra of kinetic energy (thick solid line)
and magnetic energy (thin solid line) \add{scaled up by a factor $10^2$}
for the ABC runs. The dashed 
line indicates the Kolmogorov slope as a reference.}
\end{figure}
We begin by describing the general properties of the two dynamos
investigated in this work. Figure \ref{fig_01} shows the kinetic
and magnetic energy spectra for a TG simulation in the kinematic
dynamo regime. In Figure \ref{fig_02} we show the same spectra for the
ABC run. Note that the kinetic energy spectrum peaks in both cases
at $k_F \sim 3 $, where a well-defined large scale flow is present.
For larger wavenumbers the spectrum presents a short inertial range,
with Kolmogorov scaling.
During the kinematic regime, 
the magnetic energy spectrum peaks at small scales ($k \sim 9$ for TG 
and $k \sim 12$ for ABC) and all the modes grow
exponentially with the same rate. As a result, all the spectra (and
transfer functions) preserve their dependence with wavenumber 
\add{(up to an amplitude normalization)} as time evolves.

\begin{figure*}
\includegraphics[width=16cm,height=6.5cm]{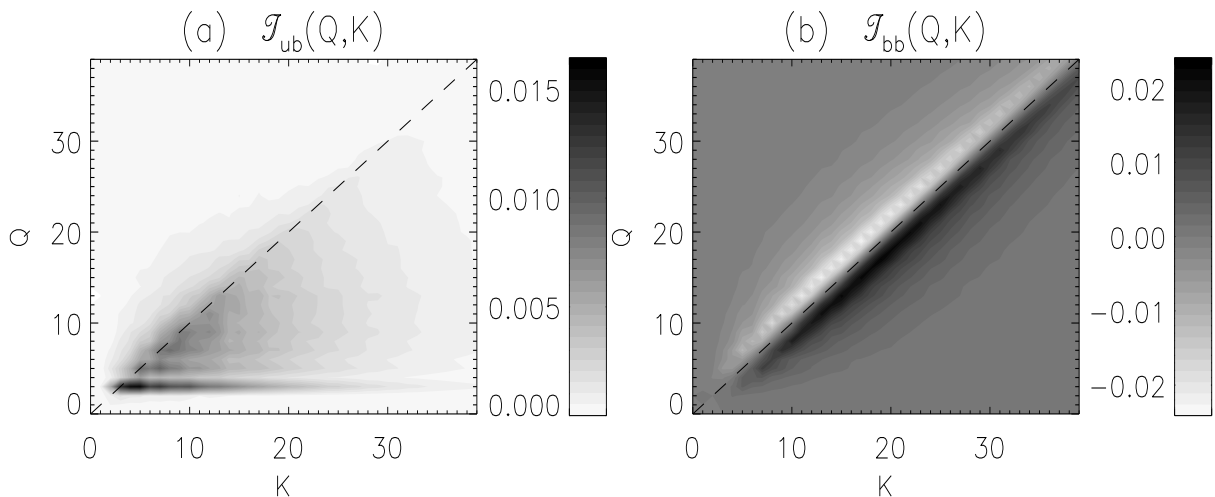}
\caption{\label{fig_03}Transfer functions (a) ${\mathcal T}_{ub}$, 
and (b) ${\mathcal T}_{bb}$ (see Eqs. (4) and (5)), as a function of $Q$ and $K$ in the TG run 
during the kinematic regime. ${\mathcal T}_{ub}$ is positive for all 
values of $Q$ and $K$ shown. \add{ Each point (Q,K) in panel (a) represents the rate of transfer
of energy from the velocity mode $Q$ to the magnetic mode $K$.
Each point (Q,K) in panel (b) represents the rate of transfer
of energy from the magnetic mode $Q$ to the magnetic mode $K$.}
The dashed line indicates the diagonal $K=Q$.}
\end{figure*}
To study the kinematic regime, the MHD simulations
were started with a random magnetic field with values of magnetic
energy as low as $E_M/E_K = 5 \times 10^{-9}$, 
to ensure that the Lorentz force was negligible at all wavenumbers 
even with the magnetic energy spectrum peaking at small scales. 

We first start with some general properties of the two transfer functions. 
Contour plots of the transfers ${\mathcal T}_{ub}(Q,K)$
and ${\mathcal T}_{bb}(Q,K)$ during the kinematic regime of the 
TG run are shown in Figure \ref{fig_03}. The gray scale indicates 
magnitude of the transfer, with `dark' being positive and `bright' 
negative. The figure should be
interpreted as follows: At a given point $(Q,K)$ on Fig. \ref{fig_03}(a),
where the transfer is positive (negative), energy is given
(received) by the velocity field at the scale $Q$ to \add{(from)} the magnetic field
at scale $K$. Similarly at a given point $(Q,K)$ on Fig. \ref{fig_03}(b),
where the transfer is positive (negative), energy is given
(received) by the magnetic field at the scale $Q$ to the magnetic field
at scale $K$. 


Note that ${\mathcal T}_{bb}$ is by definition anti-symmetric along the 
diagonal $K=Q$ and is mostly concentrated in the surroundings of the 
diagonal. The ${\mathcal T}_{ub}$ transfer is concentrated on a triangle 
below the diagonal and is positive everywhere. The fact that 
${\mathcal T}_{bb}$ is concentrated along the diagonal implies as we 
will show later locality of interactions, while the ``triangular" shape 
of ${\mathcal T}_{ub}$ implies long range interactions in Fourier space.

To draw conclusions from the functional form of the transfers we need to
examine their behavior for different fixed values of $K$ or $Q$. 
\begin{figure}
\includegraphics[width=8cm]{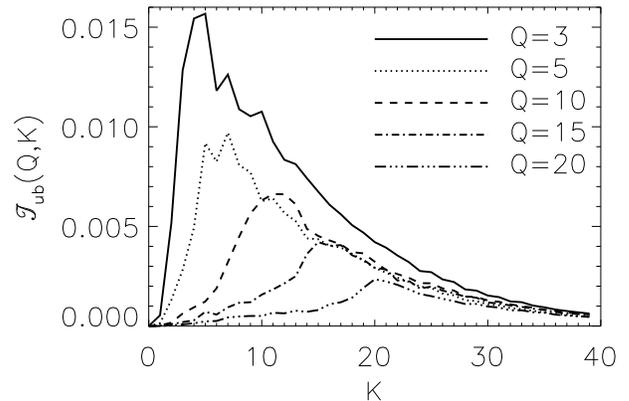}
\caption{\label{fig_04}Transfer function ${\mathcal T}_{ub}(Q,K)$ 
(from the kinetic energy at $Q$ to the magnetic energy at $K$) for fixed 
values of $Q$ during the kinematic regime of the TG run.}
\end{figure}
\begin{figure}
\includegraphics[width=8cm]{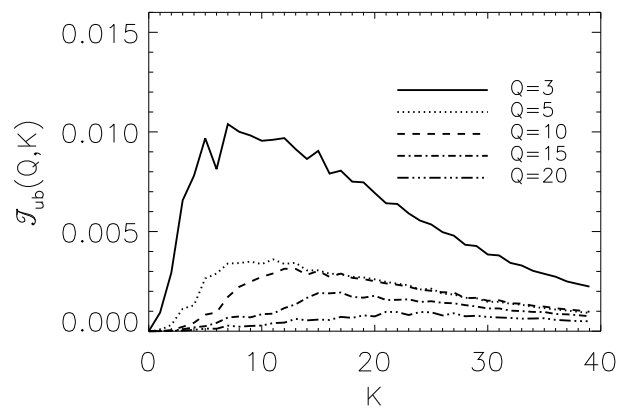}
\caption{\label{fig_05}Transfer function ${\mathcal T}_{ub}(Q,K)$ 
for fixed values of $Q$ during the kinematic regime of the ABC run.}
\end{figure}
Figures \ref{fig_04} and \ref{fig_05} show the ${\mathcal T}_{ub}(Q,K)$
function at constant values of $Q$, for the TG and ABC simulations
respectively. The transfer is always positive, implying that kinetic energy 
is transfered from all the velocity wave numbers $Q$ 
to magnetic energy at different $K$-shells. 
The transfer is maximum for wavenumbers close to $Q$, and then slowly decays. 
Note that in the ABC run the flow at $Q=3$ gives more energy than the
turbulent fluctuations ($Q=5,10,20,30$) when compared with the TG
simulation. This is related with the fact that in the ABC run the
$Q=3$ shell contains most of the kinetic helicity of the flow, an
ingredient known to be relevant for dynamo action \cite{Moffatt,Krause}.
We note here that since the transfers are normalized by the total
magnetic energy and the two runs have different magnetic energy spectra,
a direct comparison of the values of the transfers between the two runs
cannot be made.

\begin{figure}
\includegraphics[width=8cm]{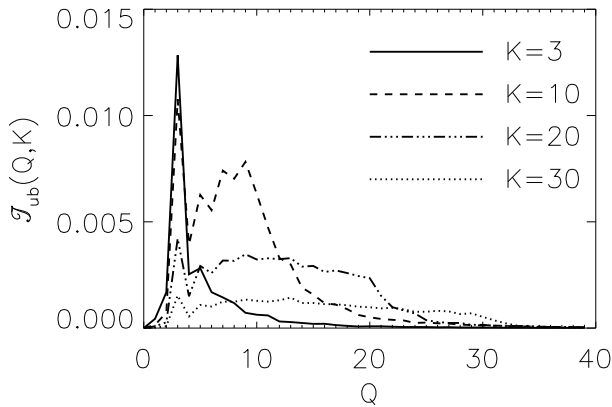}
\caption{\label{fig_06}Transfer function ${\mathcal T}_{ub}(Q,K)$ 
for fixed values of $K$ during the kinematic regime of the TG run.}
\end{figure}
\begin{figure}
\includegraphics[width=8cm]{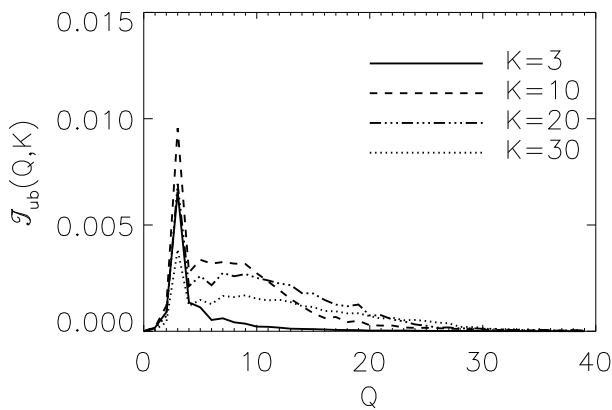}
\caption{\label{fig_07}Transfer function ${\mathcal T}_{ub}(Q,K)$ 
for fixed values of $K$ during the kinematic regime of the ABC run.}
\end{figure}
In Figures \ref{fig_06} and \ref{fig_07} we show the same transfer
function ${\mathcal T}_{ub}$ but now for constant values of $K$. 
The transfer is positive at all scales, pointing to the fact 
that all velocity
shells are giving energy to the magnetic field (compare this result
with the turbulent steady state in Paper I, where energy is being
transfered from the magnetic field to the velocity field at small
scales). A peak at $Q=3$ can be identified at all wavenumbers $K$, 
indicating that the large scale flow gives energy non-locally to all magnetic
shells. For wavenumbers $Q>3$ also a plateau can be identified, where
${\mathcal T}_{ub}$ as a function of $Q$ is approximately constant.
The plateau drops at $K \gtrsim Q$. This region of constant
${\mathcal T}_{ub}$ corresponds to all kinetic energy shells at
$3 < Q \lesssim K$ (the turbulent fluctuations) transferring the same
amount of energy to the magnetic field at the shell $K$. In the
ABC simulations, the role played by the turbulent fluctuations is
again observed to be smaller than in the TG runs when compared with
the large scale flow at $Q \sim 3$. 

\begin{figure}
\includegraphics[width=8cm]{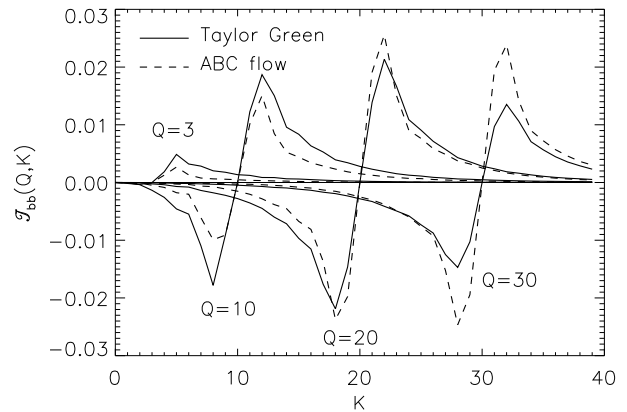}
\caption{\label{fig_08}Transfer function ${\mathcal T}_{bb}(Q,K)$ 
(from magnetic energy at shell $Q$ to magnetic energy at shell $K$) 
at fixed values of $Q$, during the kinematic regime of the TG and ABC 
runs.}
\end{figure}
The transfer of magnetic energy between different scales is shown
in Figure \ref{fig_08}. As previously mentioned, this transfer
is associated with the cascade of energy to smaller scales.
Each shell $Q$ is giving energy to a slightly
larger wavenumber $K$ (the positive peak of the curves) and receiving
energy from a slightly smaller wavenumber $K'$ (the negative peak
of the curves). There is an increase of the amplitude of the transfer
as the wavenumber $Q$ is increased up until the peak of the spectrum is
reached and then it drops again.
This transfer function drops fast for wavenumbers
$K$ and $Q$ far apart and therefore 
\add{indicates a local transfer of energy}.

\begin{figure}
\includegraphics[width=8cm]{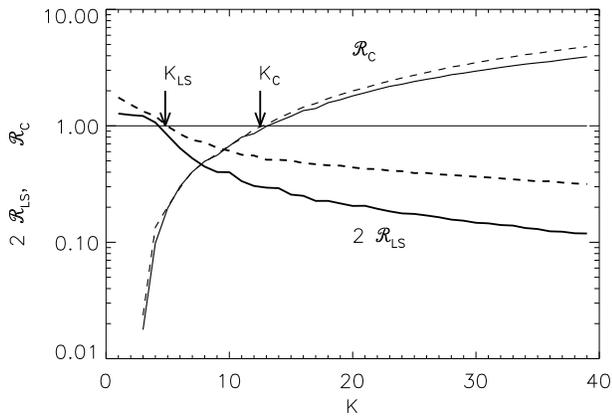}
\caption{\label{fig_09} Ratio ${\mathcal R}_{LS}$  of energy received by the magnetic field 
    at wave number $K$ from the forced wavenumbers against all the wavenumbers, 
    and ratio ${\mathcal R}_{C}$ of energy received by the magnetic 
    field at wave number $K$ from the magnetic field at larger scales 
    \add{through a cascade process}
    against 
    energy received from the velocity field. The 
    solid lines correspond to the TG run while the dashed lines correspond 
    to ABC, both in the kinematic regime.}
\end{figure}

We are ready now to answer some of the questions posed in the introduction.
First we want to consider if it is the large scale flow that drives
the dynamo or the turbulent fluctuations. On average the contribution
to the injection of magnetic energy from the large scale flow is $16\%$ 
for the TG flow, and $25\%$ for the ABC flow. Note that this fraction is 
much smaller than what is obtained in the saturated regime ($60\%$ for TG, 
and $75\%$ for ABC in \cite{Alexakis05}). Furthermore, the influence of the 
large scale flow becomes smaller as we are deeper in the inertial range. In 
Figure \ref{fig_09} we show the ratio:
\[{\mathcal R}_{LS}(K)=\sum_{Q=2,3,4}{\mathcal T}_{ub}(Q,K) / 
    \sum_Q{\mathcal T}_{ub}(Q,K) \]
that expresses the fraction of energy a magnetic shell $K$ receives only 
from the the large scale flow (the peak at $Q=2,3,4$ in Figs. 
\ref{fig_06} and \ref{fig_07}), to the total energy received by the 
same shell from the velocity field at all scales. For both flows the 
energy input from large scales becomes smaller as the wavenumber $K$ is 
increased and the large scale flow \add{only} dominates the injection of magnetic 
energy over a small range $K_F<K<K_{LS}$, with $K_{LS}\simeq 5$.


Another question we posed in the introduction is whether the small scale 
magnetic fluctuations are the result of a cascade of energy from the 
large scale magnetic field, or from a direct input of energy 
(amplification) from the velocity field. To answer this 
question, in Figure \ref{fig_09} we also plot the ratio:
\[{\mathcal R}_{C}(K)=\sum_{Q=0}^K{\mathcal T}_{bb}(Q,K) / 
    \sum_Q{\mathcal T}_{ub}(Q,K) \]
that expresses the fraction of energy a magnetic shell $K$ receives from 
the cascade of energy from larger magnetic scales to the total energy 
received in the same shell directly from the velocity field. The cascading 
term appears to be smaller up to a wavenumber
$K_C\simeq 12$ close to the peak of the magnetic energy spectrum. For 
$K>K_C$ there is more energy input from the cascade than the input from the 
velocity field. Between these two processes, a range of wavenumbers 
$K_{LS}<K<K_C$ exists where the amplification of the magnetic field
is purely dominated by injection from the turbulent velocity scales.

\begin{figure}
\includegraphics[width=8cm]{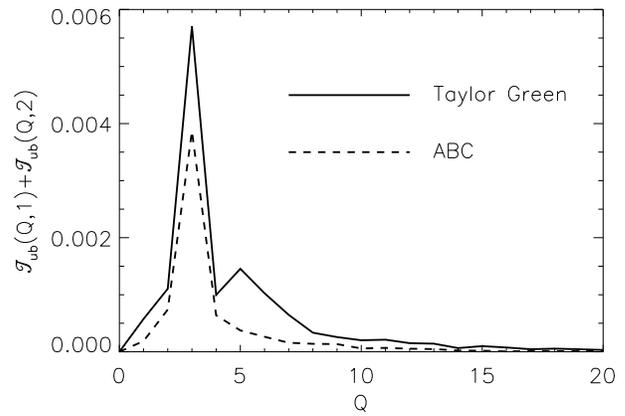}
\caption{\label{fig_10}Energy received by the magnetic field at 
    scales larger than the forcing band ($K=1$ and 2) from the velocity 
    field at wavenumbers $Q$.}
\end{figure}
We also investigate the growth rate of large scale magnetic fields restricted 
to the shells $K=1,2$. In order to 
obtain the highest possible Reynolds numbers in the simulations, the scale 
separation between the forcing band and the large scale magnetic field was chosen to be small 
 \add{and therefore an investigation of the alpha dynamo effect is not possible in the present study.
Here we just limit ourselves to investigate which scale of the velocity
field is responsible for the input of energy in the large scales $K=1$ and $K=2$
of the magnetic field.} 
In figure \ref{fig_10} we show the transfer of energy from
the velocity field to these large scale modes. Although there is a 
contribution from the turbulent fluctuations, the bulk of the energy 
originates from the forced modes. A similar result was obtained in 
Ref. \cite{Brandenburg01}, in simulations with larger scale separation 
($k_F \simeq 5$) but lower Reynolds numbers.

\begin{figure}
\includegraphics[width=8cm]{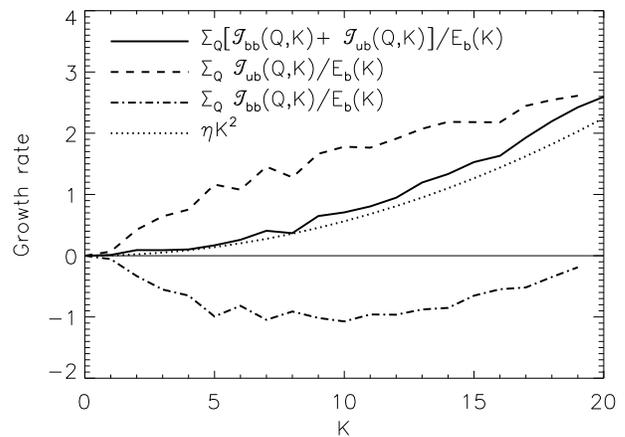}
\caption{\label{fig_11} Terms \add{determining} the growth rate of magnetic 
    energy [see Eq. (\ref{eq:growth})] as a function of wavenumber $K$ 
    for the TG run during the kinematic regime. The dashed line is 
    the energy received by the magnetic field from the velocity field, 
    and the dash-dotted line is the cascade of magnetic energy. The 
    solid line is the total energy received by the magnetic field, while 
    the dotted line is the Ohmic dissipation. The difference between the 
    \add{last} two curves gives the growth rate.} 
\end{figure}
\begin{figure}
\includegraphics[width=8cm]{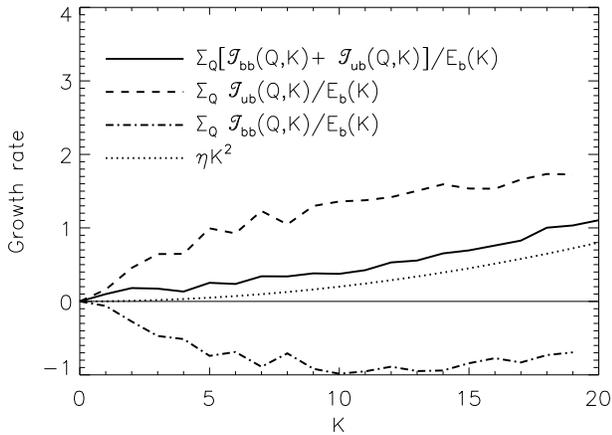}
\caption{\label{fig_12}The budget of magnetic energy giving rise to the 
    growth rate for the ABC run during the kinematic regime. Labels are 
    as in Fig. \ref{fig_11}.}
\end{figure}
Finally, in both simulations, all wavenumbers are observed to grow with
the same growth rate during the kinematic regime. 
To investigate this
we can write the energy budget using the induction equation
(\ref{eq:induc}) in Fourier space. Taking the dot product with the
magnetic field ${\bf b}_K$ at the shell $K$, and dividing by the
magnetic energy $E_b(K)$ in that shell, we finally obtain
\begin{eqnarray}
\label{growth}
\frac{1}{E_b(K)} \frac{\partial}{\partial t} E_b(K) &=&
\frac{1}{E_b(K)} \sum_Q [{\mathcal T}_{ub}(Q,K)+{\mathcal T}_{bb}(Q,K)]
    \nonumber \\
- \eta K^2, \label{eq:growth}
\end{eqnarray}
where the simplification ${\mathcal D}(K)\simeq K^2E_b(K)$ was used.
The left-hand side of equation (\ref{growth}) gives the growth rate $\sigma$. 
The first two terms in the right hand side are 
the energy received by the magnetic field at the shell $K$ from
the velocity field and from the magnetic field at all scales.
The last term is the Ohmic dissipation.
In figures \ref{fig_11} and \ref{fig_12} we show
each term of this budget as a function of the wavenumber $K$ for the
TG and ABC runs. The difference between the solid line and the dotted line 
is the
growth rate. In spite of the fluctuations, the growth rate seems to
be constant in a wide range of wavenumbers. This is more clearly
observed in the ABC run because of the larger growth rate in this
simulation. 
The constant growth rate over all scales therefore is the result
of a balance between the energy received by the magnetic field at each
shell $K$ locally (from the direct cascade), non-locally (from the
stretching of field lines), and of the Ohmic dissipation. Note that 
when integrated over all $Q$, the direct cascade ${\mathcal T}_{bb}(Q,K)$ 
gives a negative contribution (up to $k\simeq20$ in the TG run, and larger 
wavenumbers for the ABC case), indicating that each magnetic shell $K$ 
gives locally more energy to smaller scales than what it receives from 
the larger scales. This is compensated by the energy injected by the 
velocity field through the transfer ${\mathcal T}_{ub}$.

\subsection{ \label{saturation} The saturation of the dynamo }

\begin{figure}
\includegraphics[width=8cm]{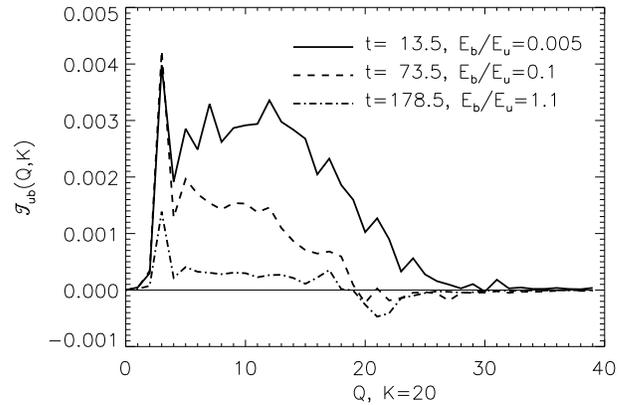}
\caption{\label{fig_13} Transfer function from the kinetic energy 
at $Q$ to the magnetic energy at $K=20$ 
\add{for three different times as the magnetic 
field approaches saturation in the TG run.}}
\end{figure}
\begin{figure}
\includegraphics[width=8cm]{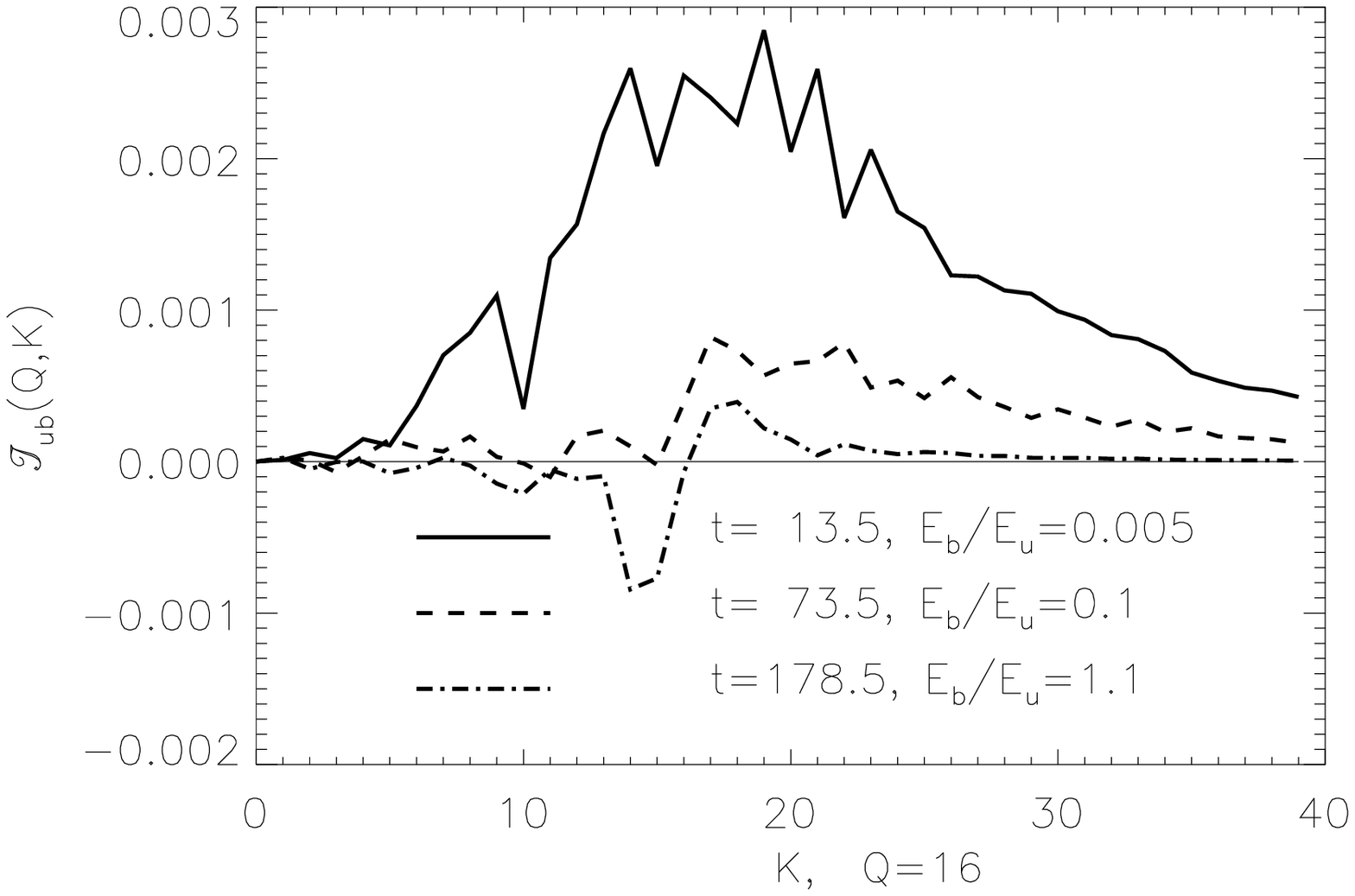}
\caption{\label{fig_14}Transfer function from the kinetic energy 
at $Q=16$ to the magnetic energy at $K$
\add{for three different times as the magnetic
field approaches saturation in the TG run.}} 
\end{figure}
\begin{figure}
\includegraphics[width=8cm]{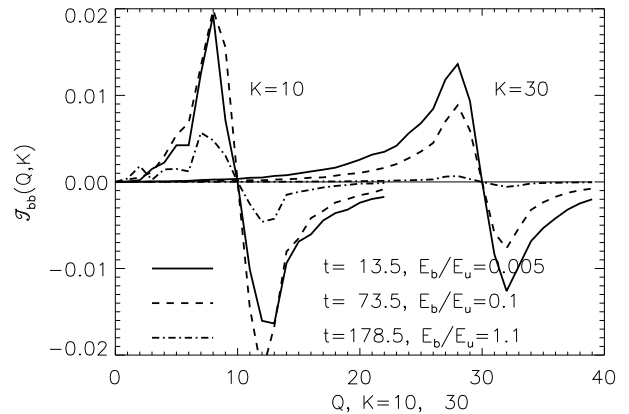}
\caption{\label{fig_15}Transfer of magnetic energy at shell $Q$ 
to shell $K$ (for $K=10$ and $K=20$)  
\add{for three different times as the magnetic
field approaches saturation in the TG run.}}
\end{figure}
In this section we discuss the evolution of the transfer function for the TG run
as the dynamo approaches the nonlinear saturation. The ABC run shows
similar features except for a slow growth of the magnetic field at $K=1$
that finally dominates the magnetic energy. The transfer of magnetic
energy at the large scales in this case has been studied in
\cite{Brandenburg01}. For details of the transfer in the final state
reached by the two simulations, we refer the reader to Paper I.

Figure \ref{fig_13} shows the ${\mathcal T}_{ub}$ transfer at $K=20$
as a function of time, as the nonlinear saturation takes place. Each
transfer has been normalized by the total magnetic energy at that time.
The transfer at $t=13.5$ corresponds to the kinematic regime. At $t=73.5$,
the small scale magnetic field saturates and stops growing (see
\cite{Mininni05a}). The velocity field turbulent fluctuations are
\add{partially} quenched, and the kinetic energy at small scales is reduced. This
suppression of turbulence by the magnetic field has been previously
observed \cite{Brandenburg01,Mininni05a}, 
and as a result the transfer of energy 
to the magnetic field at wavenumber $K$
from the velocity field between $3<Q \lesssim K$
is also strongly reduced. However, the large scale velocity field at
$Q=3$ keeps transferring energy to the magnetic field. In this stage,
the large scale magnetic field keeps growing until the large scales
are dominated by magnetic energy, and suppress even more the turbulent
fluctuations. At $t=178.5$, the system has finally reached the steady
state. The magnetic field at each shell $K$ is sustained by both the
large scale flow and the turbulent fluctuations, but now the amplitude
at all scales has been \add{reduced due to}  the Lorentz force. Note also that
the transfer at scales $Q \gtrsim K$ is now negative, pointing out to the fact that
the magnetic energy is feeding the velocity field at small scales.

In figure \ref{fig_14}, we show the transfer ${\mathcal T}_{ub}(Q,K)$
for a fixed velocity wave number
$Q=16$ for the same times as in figure \ref{fig_13}. Again,
the transfers have been normalized by the total magnetic energy at
each time. In the kinematic regime, the velocity field at $Q=16$ gives
energy (positive transfer) to the magnetic field at all scales. When
the small scale magnetic field saturates ($t=73.5$) the velocity field
at $Q=16$ gives energy only to magnetic shells with $K \ge Q$, and modes
with $K < Q$ receive almost no energy. This regime corresponds to the
case where the magnetic field at small scales has saturated and its energy
is sustained by the dynamo without further amplification. The large
scale field keeps growing, mostly fed by the large scale flow
at $Q=3$ as previously discussed. Finally, in the saturated regime
($t=178.5$) the magnetic field at $Q=16$ receives energy from all the
kinetic shells $Q$ with $K \ge Q$, while it gives energy (negative transfer)
to all kinetic shells with $K < Q$.

Finally, in
Figure \ref{fig_15} we show the transfer of magnetic to magnetic
energy ${\mathcal T}_{bb}$ as a function of time. The local transfer
between magnetic shells is not changing as much as ${\mathcal T}_{ub}$
as the dynamo saturates, except for a change in the amplitude. 
\add{The amplitude is decreasing as saturation is approached,
an effect that appears first in the small scales.
We note that the role of the local direct cascade ${\mathcal T}_{bb}$
becomes more dominant when compared with the ${\mathcal T}_{bb}$ term
as saturation is approached \cite{Alexakis05}.}

\section{ \label{Discussion} Discussion }

In this work, the transfer of energy from the different scales of the velocity field
to the different scales of the magnetic field has been studied.
We now give a brief summary of our most important results, and discuss
the implications for dynamo theory.

It has been shown that the magnetic field grows as the result of a complex interaction
between large and small scales. Both large and small scales of the flow give a
contribution to the dynamo. 
The amplification of the large scale magnetic field 
during the kinematic regime is due to
the large scale flow (large scale dynamo).
At smaller scales, most of the
injection of energy from the velocity field 
during the kinematic regime is due to the turbulent
fluctuations of the velocity field (small scale dynamo). 
A competing mechanism for the amplification of  
the magnetic field at the small scales is the cascade of magnetic field energy
from the large scales to the small scales that is also transferring energy
to the small scales. 
The rate that energy is transfered to the small scales through the
cascading process is smaller than the rate that the velocity field is 
injecting energy at the small scales for a finite range of wave numbers.  
For sufficiently small scales 
(close to the scale at which the peak of the magnetic energy spectrum is reached), 
the cascading term becomes larger than the small scale dynamo term.
 
The results in this paper and the formalism used
help us understand and classify the dynamo processes involved
in the amplification of the magnetic field. In this formalism,
we were able to measure and compare each component in the dynamo process that
is involved in the amplification of the magnetic field in both small 
and large scales. Therefore, 
we can distinguish
between different dynamos based on whether the cascading terms dominate over
the injection terms in the small scales and on whether the turbulent fluctuations
are more dominant for the generation of the magnetic field when compared with the
input from the large scale flow. Of course, this
is not the only possible distinction that can be made between dynamos; 
however, it is an important step 
towards classifying dynamos in the presence of both a large scale
flow and turbulent fluctuations.

Finally we would like to note that the investigation of the growth of large scale magnetic
field with enough scale separation between  the forced scale and the domain
size is required to study \add{processes such as the dynamo $\alpha$-effect 
and the inverse cascade of magnetic helicity}.
A similar analysis \add{will be performed in this context in future works.}


\begin{acknowledgments}
The authors are grateful to J. Herring for valuable discussions
and his careful reading of this document.
Computer time was provided by NCAR. The NSF grant CMG-0327888
at NCAR supported this work in part and is gratefully acknowledged.
\end{acknowledgments}

\bibliography{ms}

\begin{thebibliography}{26}
\expandafter\ifx\csname natexlab\endcsname\relax\def\natexlab#1{#1}\fi
\expandafter\ifx\csname bibnamefont\endcsname\relax
  \def\bibnamefont#1{#1}\fi
\expandafter\ifx\csname bibfnamefont\endcsname\relax
  \def\bibfnamefont#1{#1}\fi
\expandafter\ifx\csname citenamefont\endcsname\relax
  \def\citenamefont#1{#1}\fi
\expandafter\ifx\csname url\endcsname\relax
  \def\url#1{\texttt{#1}}\fi
\expandafter\ifx\csname urlprefix\endcsname\relax\def\urlprefix{URL }\fi
\providecommand{\bibinfo}[2]{#2}
\providecommand{\eprint}[2][]{\url{#2}}

\bibitem[{\citenamefont{Moffatt}(1978)}]{Moffatt}
\bibinfo{author}{\bibfnamefont{H.~K.} \bibnamefont{Moffatt}},
  \emph{\bibinfo{title}{Magnetic field generation in electrically conducting
  fluids}} (\bibinfo{publisher}{Cambridge Univ. Press},
  \bibinfo{address}{Cambridge}, \bibinfo{year}{1978}).

\bibitem[{\citenamefont{Dikpati and Charbonneau}(1999)}]{Dikpati99}
\bibinfo{author}{\bibfnamefont{M.}~\bibnamefont{Dikpati}} \bibnamefont{and}
  \bibinfo{author}{\bibfnamefont{P.}~\bibnamefont{Charbonneau}},
  \bibinfo{journal}{Astrophys.\ J.} \textbf{\bibinfo{volume}{518}},
  \bibinfo{pages}{508} (\bibinfo{year}{1999}).

\bibitem[{\citenamefont{Cattaneo}(1999)}]{Cattaneo99}
\bibinfo{author}{\bibfnamefont{F.}~\bibnamefont{Cattaneo}},
  \bibinfo{journal}{Astrophys.\ J.} \textbf{\bibinfo{volume}{515}},
  \bibinfo{pages}{L39} (\bibinfo{year}{1999}).

\bibitem[{\citenamefont{Vainshtein and Zeldovich}(1972)}]{Vainshtein72}
\bibinfo{author}{\bibfnamefont{S.~I.} \bibnamefont{Vainshtein}}
  \bibnamefont{and} \bibinfo{author}{\bibfnamefont{Y.~B.}
  \bibnamefont{Zeldovich}}, \bibinfo{journal}{Sov.\ Phys.\ Usp.}
  \textbf{\bibinfo{volume}{15}}, \bibinfo{pages}{159} (\bibinfo{year}{1972}).

\bibitem[{\citenamefont{Steenbeck et~al.}(1966)\citenamefont{Steenbeck, Krause,
  and Raedler}}]{Steenbeck66}
\bibinfo{author}{\bibfnamefont{M.}~\bibnamefont{Steenbeck}},
  \bibinfo{author}{\bibfnamefont{F.}~\bibnamefont{Krause}}, \bibnamefont{and}
  \bibinfo{author}{\bibfnamefont{K.-H.} \bibnamefont{Raedler}},
  \bibinfo{journal}{Z.\ Naturforsch.} \textbf{\bibinfo{volume}{21a}},
  \bibinfo{pages}{369} (\bibinfo{year}{1966}).

\bibitem[{\citenamefont{Krause and Raedler}(1980)}]{Krause}
\bibinfo{author}{\bibfnamefont{F.}~\bibnamefont{Krause}} \bibnamefont{and}
  \bibinfo{author}{\bibfnamefont{K.-H.} \bibnamefont{Raedler}},
  \emph{\bibinfo{title}{Mean-field magnetohydrodynamics and dynamo theory}}
  (\bibinfo{publisher}{Pergamon Press}, \bibinfo{address}{New York},
  \bibinfo{year}{1980}).

\bibitem[{\citenamefont{Pouquet et~al.}(1976)\citenamefont{Pouquet, Frisch, and
  L\'eorat}}]{Pouquet76}
\bibinfo{author}{\bibfnamefont{A.}~\bibnamefont{Pouquet}},
  \bibinfo{author}{\bibfnamefont{U.}~\bibnamefont{Frisch}}, \bibnamefont{and}
  \bibinfo{author}{\bibfnamefont{J.}~\bibnamefont{L\'eorat}},
  \bibinfo{journal}{J.\ Fluid Mech.} \textbf{\bibinfo{volume}{77}},
  \bibinfo{pages}{321} (\bibinfo{year}{1976}).

\bibitem[{\citenamefont{Meneguzzi et~al.}(1981)\citenamefont{Meneguzzi, Frisch,
  and Pouquet}}]{Meneguzzi81}
\bibinfo{author}{\bibfnamefont{M.}~\bibnamefont{Meneguzzi}},
  \bibinfo{author}{\bibfnamefont{U.}~\bibnamefont{Frisch}}, \bibnamefont{and}
  \bibinfo{author}{\bibfnamefont{A.}~\bibnamefont{Pouquet}},
  \bibinfo{journal}{Phys.\ Rev.\ Lett.} \textbf{\bibinfo{volume}{47}},
  \bibinfo{pages}{1060} (\bibinfo{year}{1981}).

\bibitem[{\citenamefont{Brandenburg}(2001)}]{Brandenburg01}
\bibinfo{author}{\bibfnamefont{A.}~\bibnamefont{Brandenburg}},
  \bibinfo{journal}{Astrophys.\ J.} \textbf{\bibinfo{volume}{550}},
  \bibinfo{pages}{824} (\bibinfo{year}{2001}).

\bibitem[{\citenamefont{G\'omez and Mininni}(2004)}]{Gomez04}
\bibinfo{author}{\bibfnamefont{D.~O.} \bibnamefont{G\'omez}} \bibnamefont{and}
  \bibinfo{author}{\bibfnamefont{P.~D.} \bibnamefont{Mininni}},
  \bibinfo{journal}{Nonlin.\ Proc.\ Geophys.} \textbf{\bibinfo{volume}{11}},
  \bibinfo{pages}{619} (\bibinfo{year}{2004}).

\bibitem[{\citenamefont{Gilbert et~al.}(1988)\citenamefont{Gilbert, Frisch, and
  Pouquet}}]{Gilbert88}
\bibinfo{author}{\bibfnamefont{A.}~\bibnamefont{Gilbert}},
  \bibinfo{author}{\bibfnamefont{U.}~\bibnamefont{Frisch}}, \bibnamefont{and}
  \bibinfo{author}{\bibfnamefont{A.}~\bibnamefont{Pouquet}},
  \bibinfo{journal}{Geophys. Astrophys. Fluid Dynamics}
  \textbf{\bibinfo{volume}{42}}, \bibinfo{pages}{151} (\bibinfo{year}{1988}).

\bibitem[{\citenamefont{Nore et~al.}(1997)\citenamefont{Nore, Brachet,
  Politano, and Pouquet}}]{Nore97}
\bibinfo{author}{\bibfnamefont{C.}~\bibnamefont{Nore}},
  \bibinfo{author}{\bibfnamefont{M.~E.} \bibnamefont{Brachet}},
  \bibinfo{author}{\bibfnamefont{H.}~\bibnamefont{Politano}}, \bibnamefont{and}
  \bibinfo{author}{\bibfnamefont{A.}~\bibnamefont{Pouquet}},
  \bibinfo{journal}{Phys.\ Plasmas Lett.} \textbf{\bibinfo{volume}{4}},
  \bibinfo{pages}{1} (\bibinfo{year}{1997}).

\bibitem[{\citenamefont{Urpin}(2002)}]{Urpin02}
\bibinfo{author}{\bibfnamefont{V.}~\bibnamefont{Urpin}},
  \bibinfo{journal}{Phys.\ Rev.\ E} \textbf{\bibinfo{volume}{65}},
  \bibinfo{pages}{026301} (\bibinfo{year}{2002}).

\bibitem[{\citenamefont{Geppert and Rheinhardt}(2002)}]{Geppert02}
\bibinfo{author}{\bibfnamefont{U.}~\bibnamefont{Geppert}} \bibnamefont{and}
  \bibinfo{author}{\bibfnamefont{M.}~\bibnamefont{Rheinhardt}},
  \bibinfo{journal}{Astron.\ Astrophys.} \textbf{\bibinfo{volume}{392}},
  \bibinfo{pages}{1015} (\bibinfo{year}{2002}).

\bibitem[{\citenamefont{Kazanstev}(1968)}]{Kazantsev68}
\bibinfo{author}{\bibfnamefont{A.~P.} \bibnamefont{Kazanstev}},
  \bibinfo{journal}{Sov.\ Phys.\ JETP} \textbf{\bibinfo{volume}{26}},
  \bibinfo{pages}{1031} (\bibinfo{year}{1968}).

\bibitem[{\citenamefont{Zeldovich et~al.}(1983)\citenamefont{Zeldovich,
  Ruzmaikin, and Sokoloff}}]{Zeldovich}
\bibinfo{author}{\bibfnamefont{Y.~B.} \bibnamefont{Zeldovich}},
  \bibinfo{author}{\bibfnamefont{A.~A.} \bibnamefont{Ruzmaikin}},
  \bibnamefont{and} \bibinfo{author}{\bibfnamefont{D.~D.}
  \bibnamefont{Sokoloff}}, \emph{\bibinfo{title}{Magnetic fields in
  astrophysics}} (\bibinfo{publisher}{Gordon and Breach Science Pub.},
  \bibinfo{address}{New York}, \bibinfo{year}{1983}).

\bibitem[{\citenamefont{Schekochihin et~al.}(2002)\citenamefont{Schekochihin,
  Boldyrev, and Kulsrud}}]{Schekochihin02}
\bibinfo{author}{\bibfnamefont{A.~A.} \bibnamefont{Schekochihin}},
  \bibinfo{author}{\bibfnamefont{S.}~\bibnamefont{Boldyrev}}, \bibnamefont{and}
  \bibinfo{author}{\bibfnamefont{R.~M.} \bibnamefont{Kulsrud}},
  \bibinfo{journal}{Astrophys.\ J.} \textbf{\bibinfo{volume}{567}},
  \bibinfo{pages}{828} (\bibinfo{year}{2002}).

\bibitem[{\citenamefont{Haugen et~al.}(2003)\citenamefont{Haugen, Brandenburg,
  and Dobler}}]{Haugen03}
\bibinfo{author}{\bibfnamefont{N.~E.~L.} \bibnamefont{Haugen}},
  \bibinfo{author}{\bibfnamefont{A.}~\bibnamefont{Brandenburg}},
  \bibnamefont{and} \bibinfo{author}{\bibfnamefont{W.}~\bibnamefont{Dobler}},
  \bibinfo{journal}{Astrophys.\ J.} \textbf{\bibinfo{volume}{597}},
  \bibinfo{pages}{L141} (\bibinfo{year}{2003}).

\bibitem[{\citenamefont{Schekochihin
  et~al.}(2004{\natexlab{a}})\citenamefont{Schekochihin, Cowley, Taylor, Maron,
  and McWilliams}}]{Schekochihin04b}
\bibinfo{author}{\bibfnamefont{A.~A.} \bibnamefont{Schekochihin}},
  \bibinfo{author}{\bibfnamefont{S.~C.} \bibnamefont{Cowley}},
  \bibinfo{author}{\bibfnamefont{S.~F.} \bibnamefont{Taylor}},
  \bibinfo{author}{\bibfnamefont{J.~L.} \bibnamefont{Maron}}, \bibnamefont{and}
  \bibinfo{author}{\bibfnamefont{J.~C.} \bibnamefont{McWilliams}},
  \bibinfo{journal}{Astrophys.\ J.} \textbf{\bibinfo{volume}{612}},
  \bibinfo{pages}{276} (\bibinfo{year}{2004}{\natexlab{a}}).

\bibitem[{\citenamefont{Batchelor}(1950)}]{Batchelor50}
\bibinfo{author}{\bibfnamefont{G.~K.} \bibnamefont{Batchelor}},
  \bibinfo{journal}{Proc.\ R.\ Soc.\ Lond.\ A} \textbf{\bibinfo{volume}{201}},
  \bibinfo{pages}{405} (\bibinfo{year}{1950}).

\bibitem[{\citenamefont{Schekochihin
  et~al.}(2004{\natexlab{b}})\citenamefont{Schekochihin, Cowley, Maron, and
  McWilliams}}]{Schekochihin04a}
\bibinfo{author}{\bibfnamefont{A.~A.} \bibnamefont{Schekochihin}},
  \bibinfo{author}{\bibfnamefont{S.~C.} \bibnamefont{Cowley}},
  \bibinfo{author}{\bibfnamefont{J.~L.} \bibnamefont{Maron}}, \bibnamefont{and}
  \bibinfo{author}{\bibfnamefont{J.~C.} \bibnamefont{McWilliams}},
  \bibinfo{journal}{Phys.\ Rev.\ Lett.} \textbf{\bibinfo{volume}{92}},
  \bibinfo{pages}{054502} (\bibinfo{year}{2004}{\natexlab{b}}).

\bibitem[{\citenamefont{Ponty et~al.}(2005)\citenamefont{Ponty, Mininni,
  Montgomery, Pinton, Politano, and Pouquet}}]{Ponty05}
\bibinfo{author}{\bibfnamefont{Y.}~\bibnamefont{Ponty}},
  \bibinfo{author}{\bibfnamefont{P.~D.} \bibnamefont{Mininni}},
  \bibinfo{author}{\bibfnamefont{D.~C.} \bibnamefont{Montgomery}},
  \bibinfo{author}{\bibfnamefont{J.-F.} \bibnamefont{Pinton}},
  \bibinfo{author}{\bibfnamefont{H.}~\bibnamefont{Politano}}, \bibnamefont{and}
  \bibinfo{author}{\bibfnamefont{A.}~\bibnamefont{Pouquet}},
  \bibinfo{journal}{Phys.\ Rev.\ Lett.} \textbf{\bibinfo{volume}{94}},
  \bibinfo{pages}{164502} (\bibinfo{year}{2005}).

\bibitem[{\citenamefont{Subramanian}(1999)}]{Subramanian99}
\bibinfo{author}{\bibfnamefont{K.}~\bibnamefont{Subramanian}},
  \bibinfo{journal}{Phys.\ Rev.\ Lett.} \textbf{\bibinfo{volume}{83}},
  \bibinfo{pages}{2957} (\bibinfo{year}{1999}).

\bibitem[{\citenamefont{Alexakis et~al.}(2005)\citenamefont{Alexakis, Mininni,
  and Pouquet}}]{Alexakis05}
\bibinfo{author}{\bibfnamefont{A.}~\bibnamefont{Alexakis}},
  \bibinfo{author}{\bibfnamefont{P.~D.} \bibnamefont{Mininni}},
  \bibnamefont{and} \bibinfo{author}{\bibfnamefont{A.}~\bibnamefont{Pouquet}},
  \bibinfo{journal}{Phys.\ Rev.\ E}  (\bibinfo{year}{2005}),
  \bibinfo{note}{submitted}.

\bibitem[{\citenamefont{Mininni
  et~al.}(2005{\natexlab{a}})\citenamefont{Mininni, Ponty, Montgomery,
  J.-F.Pinton, Politano, and Pouquet}}]{Mininni05a}
\bibinfo{author}{\bibfnamefont{P.~D.} \bibnamefont{Mininni}},
  \bibinfo{author}{\bibfnamefont{Y.}~\bibnamefont{Ponty}},
  \bibinfo{author}{\bibfnamefont{D.~C.} \bibnamefont{Montgomery}},
  \bibinfo{author}{\bibnamefont{J.-F.Pinton}},
  \bibinfo{author}{\bibfnamefont{H.}~\bibnamefont{Politano}}, \bibnamefont{and}
  \bibinfo{author}{\bibfnamefont{A.}~\bibnamefont{Pouquet}},
  \bibinfo{journal}{Astrophys.\ J.}  (\bibinfo{year}{2005}{\natexlab{a}}),
  \bibinfo{note}{in press}, \eprint{astro-ph/0412071}.

\bibitem[{\citenamefont{Mininni
  et~al.}(2005{\natexlab{b}})\citenamefont{Mininni, Montgomery, and
  Pouquet}}]{Mininni05b}
\bibinfo{author}{\bibfnamefont{P.~D.} \bibnamefont{Mininni}},
  \bibinfo{author}{\bibfnamefont{D.~C.} \bibnamefont{Montgomery}},
  \bibnamefont{and} \bibinfo{author}{\bibfnamefont{A.}~\bibnamefont{Pouquet}},
  \bibinfo{journal}{Phys.\ Rev.\ E} \textbf{\bibinfo{volume}{71}},
  \bibinfo{pages}{046304} (\bibinfo{year}{2005}{\natexlab{b}}).

\end{thebibliography}

\end{document}